\documentclass[twoside]{sfm}

\input{sf.def}

\begin{document}

\def\llm{{\sc LLmodels}}
\def\atl{{\sc ATLAS9}}
\def\aatl{{\sc ATLAS12}}
\def\starsp{{\sc STARSP}}
\def\aur{$\Theta$~Aur}
\def\logg{\log g}
\def\tauros{\tau_{\rm Ross}}
\def\kms{km\,s$^{-1}$}
\def\bz{$\langle B_{\rm z} \rangle$}
\def\degr{^\circ}
\def\aaps{A\&AS}
\def\aap{A\&A}
\def\apjs{ApJS}
\def\apj{ApJ}
\def\rmxaa{Rev. Mexicana Astron. Astrofis.}
\def\mnras{MNRAS}
\def\actaa{Acta Astron.}
\newcommand{\Tef}{T$_{\rm eff}$~}
\newcommand{\Vt}{$V_t$}
\newcommand{\CC}{$^{12}$C/$^{13}$C~}
\newcommand{\CDC}{$^{12}$C/$^{13}$C~}

\pagebreak

\thispagestyle{titlehead}

\setcounter{section}{0}
\setcounter{figure}{0}
\setcounter{table}{0}

\titl{Observed variations of the global longitudinal magnetic fields of stars.}{Bychkov V.D.$^{1}$, Bychkova L.V.$^{1}$, Madej J.$^{2}$}
{$^{1}$Special Astrophysical Observatory RAS, Russia \\
     $^{2}$ Warsaw University Observatory, Poland }

\abstre{Actually there exist 218 stars with measured phase curves of the longitudinal (effective) magnetic field $B_{e}$. 
In that group 172 objects are classified as magnetic chemically peculiar stars (mCP).
Remaining objects are stars of various spectral types, from the most massive hot Of? supergiants
to low mass red dwarfs and stars with planets. This paper aims to review briefly properties of the observed magnetic field
in various types of stars.}

\baselineskip 12pt

\section{Introduction}
Probably all stars developed global magnetic fields, and their longitudinal
field $B_e$ is variable (again probably) mostly with the period of rotation.
Strength of global fields generally is not high and yet could not be
measured in most objects. At present we observe
rapid progress in the field of instrumentation and methods of magnetic 
field measurements. All those improvements
allowed one to investigate and observe variability of magnetic fields in
Of? stars which are very difficult to measure Wade et al. \cite{Wade_2011},\cite{Wade_2012}.
Investigation of Of? objects is extremely important for the study of the
origin of global magnetic fields. \\
We list here the most obvious advantages of the above progress:\\
1. There is accumulated a large set of $B_e$ measurements.\\
2. In some cases new magnetic measurements were obtained with spectra
       of relatively low resolution.\\
3. Those data were accumulated during a long time period (over 60 years),
       which actually allows one to study the long-period magnetic behavior
       in some objects. \\
Some stars were simultaneously put into two different classes. For example, HD 96446 belongs to He-r 
and $\beta$ Cep classes and HD 97048 belongs to
TTS and Ae/Be Herbig types. Binary system DT Vir consists of two companions: UV+RS (Flare + RS CVn type stars). 
Therefore, the distribution of stars between classes had to be arbitrary or not unique in some cases.

\begin{table}[!t]
\centering
\caption{Number of objects for which we determined magnetic phase curves vs. the most important types. }
\label{list of stars} \tabcolsep1.2mm
\begin{tabular}{|l|r|}
\hline
All stars with mag. phase curves   &  218  \\
\hline
mCP stars                &  172  \\
\hline
Ae/Be Herbig stars      &    7  \\
\hline 
Be stars                 &    7  \\
\hline 
Super massive Of?           &    3  \\
\hline 
Normal early B stars        &    5  \\
\hline
Flare stars                 &    3  \\ 
\hline 
TTS (T Tau type)            &    2  \\ 
\hline 
var. Beta Cep type          &    6  \\
\hline 
SPBS                        &    3  \\
\hline 
var.BY Dra type     &    4  \\ 
\hline 
var.RS CVn type     &    1  \\ 
\hline 
Semi-reguliar var.  &    1  \\ 
\hline 
DA                  &    1  \\
\hline 
var.pulsating stars &    2  \\ 
\hline 
HPMS (high proper motions stars)      &    3  \\ 
\hline
var.Ori type        &    2  \\
\hline
\end{tabular}
\end{table}

Fig. 1 shows the discrete distribution of star number vs. spectral class in our
sample of magnetized stars with well known phase curves. Dominant fraction of
those stars is located in the spectral range from early B to F type, i.e. in the
spectral region containing mCP stars.

At present direct observations of the longitudinal magnetic field were 
obtained for ca. 1900 stars.
Only for 218 stars the number of available observations allowed us to determine
magnetic phase curves (MPC) and their parameters.

Actually we prepare a new extended catalog of the magnetic phase curves for
stars on basis of the above data.

\begin{figure}[!t]
\begin{center}
\hbox{
\includegraphics[width=4cm]{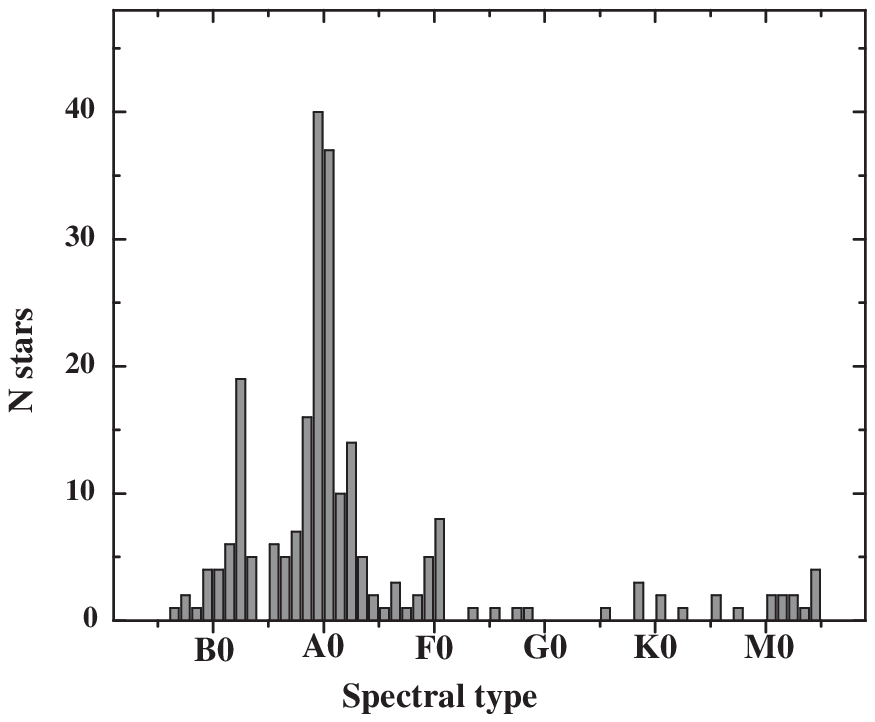}
\includegraphics[width=4cm]{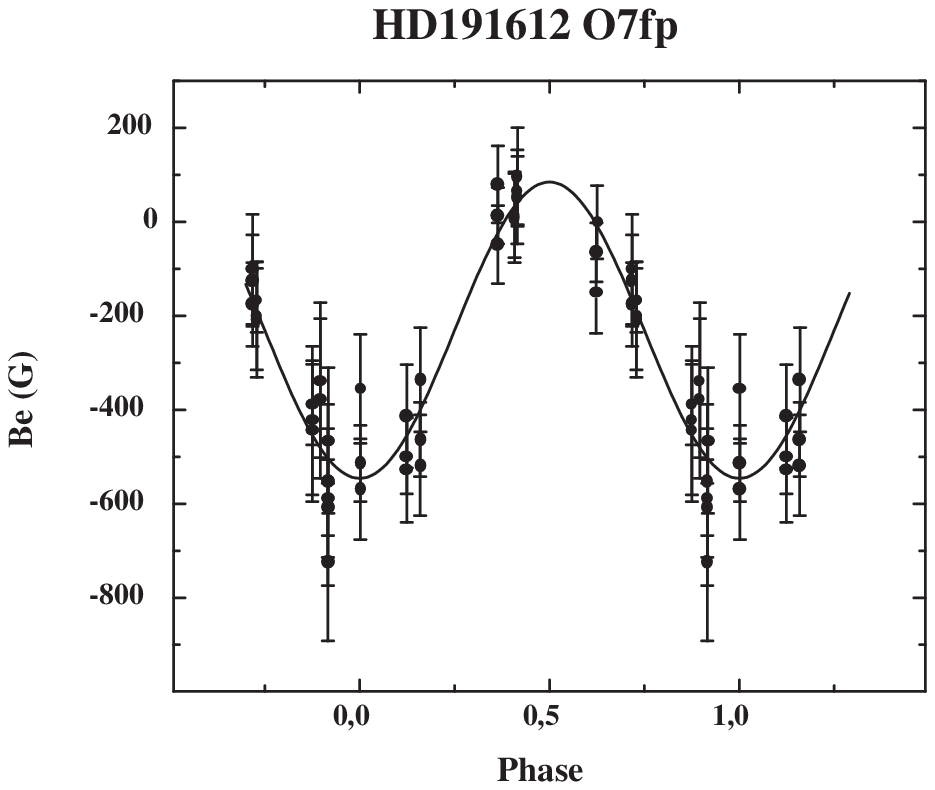}
\includegraphics[width=4cm]{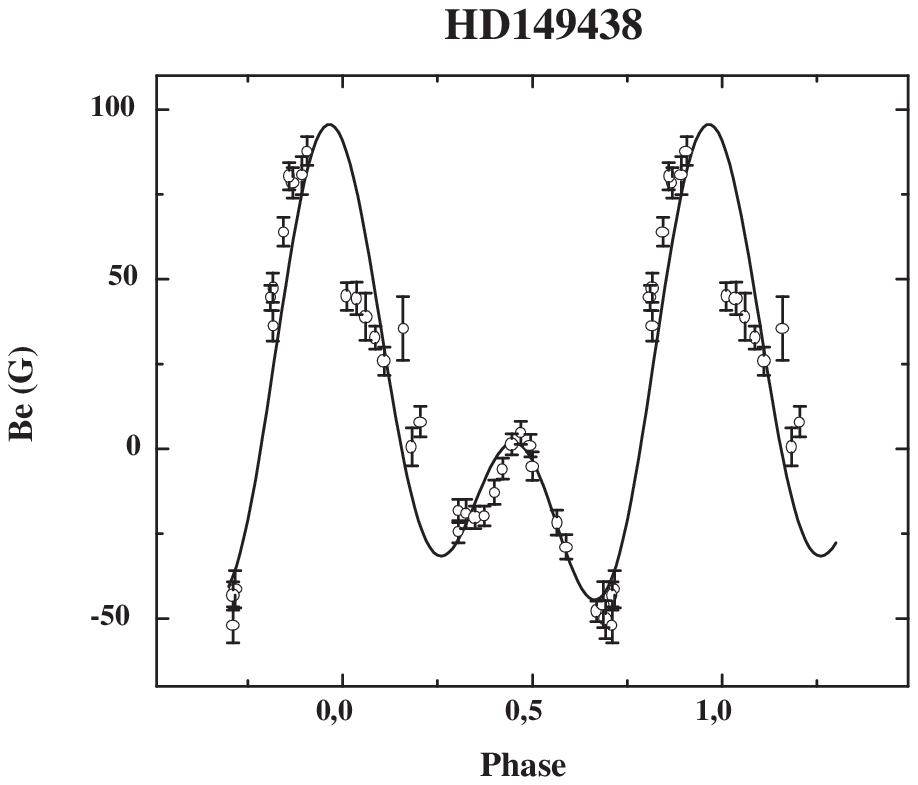}}
\vspace{-5mm}
\caption{ Figure shows the number distribution of stars with known $B_{e}(\phi)$ phase curves vs. spectral type (left panel),
magnetic phase curves of the Of?p star HD191612 with the rotational period $537^{d}.6$ following Wade et al. \cite{Wade_2011} (center panel),
for normal B star HD149438 with the rotational period $41.033$ days following Donati et al, \cite{Don_2006} (right panel) }
\label{fig:1}
\end{center}
\end{figure}

\begin{figure}[!t]
\begin{center}
\hbox{
  \includegraphics[width=4cm]{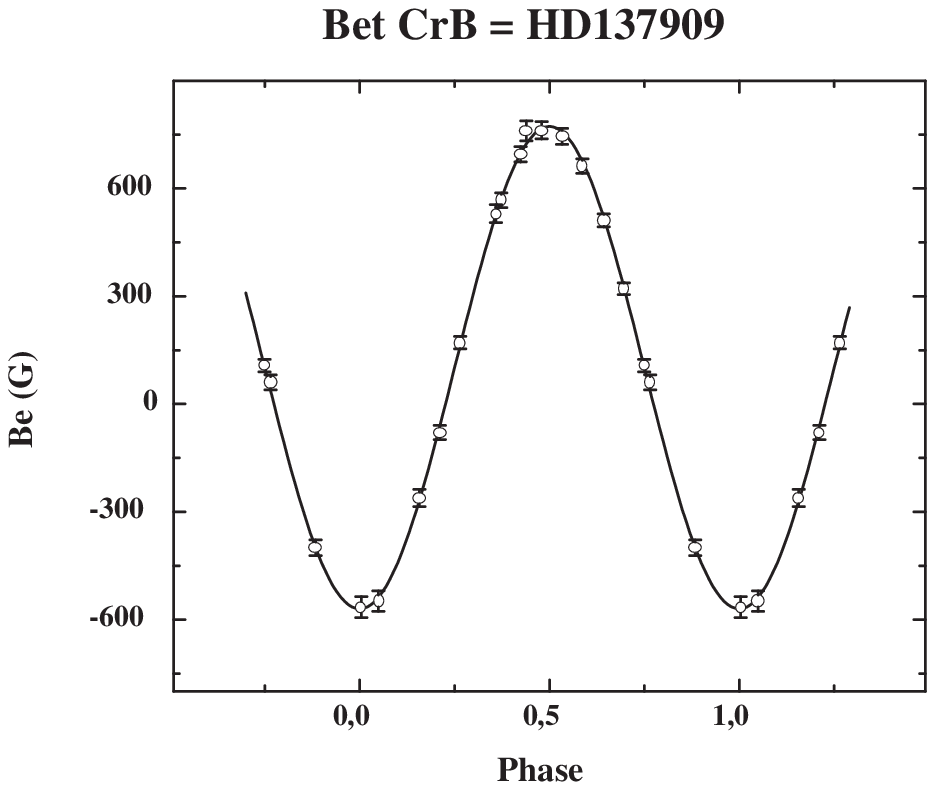}
  \includegraphics[width=4cm]{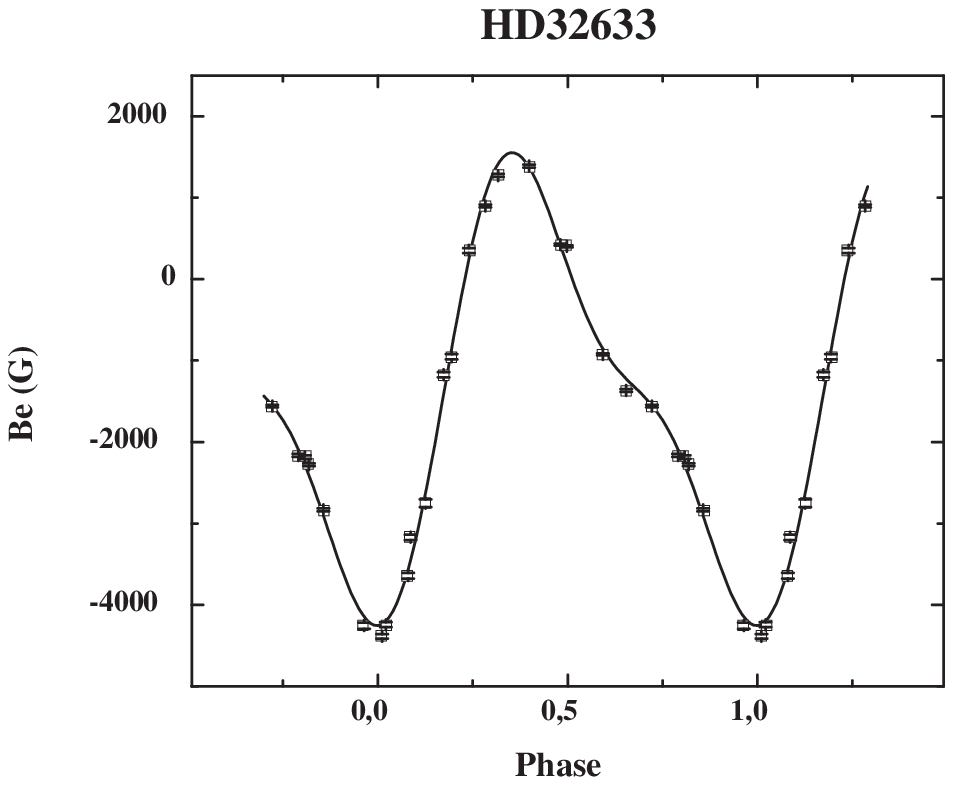}
  \includegraphics[width=4cm]{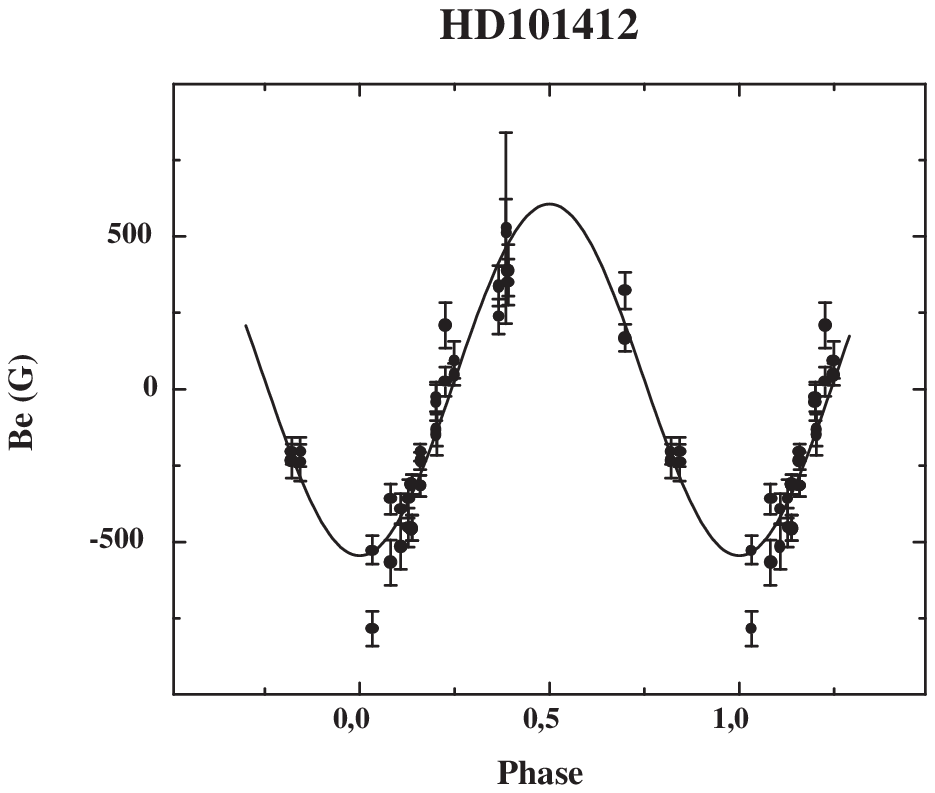}}
\vspace{-5mm}
\caption[]{
Magnetic phase curves for for the mCP-star $\beta$ CrB (HD 137909) for the high accuracy rotational period by Wade et al. \cite{Wade_2000} (left panel),
for the mCP star HD 32633 obtained from high accuracy observations by Silvester et al. \cite{Silves_2012} (center panel), 
and Ae/Be Herbig star HD101412 with the rotational period $42^{d}.076$ (right panel).}
\label{fig_box1}
\end{center}
\end{figure}

\begin{figure}[!t]
\begin{center}
\hbox{
  \includegraphics[width=4cm]{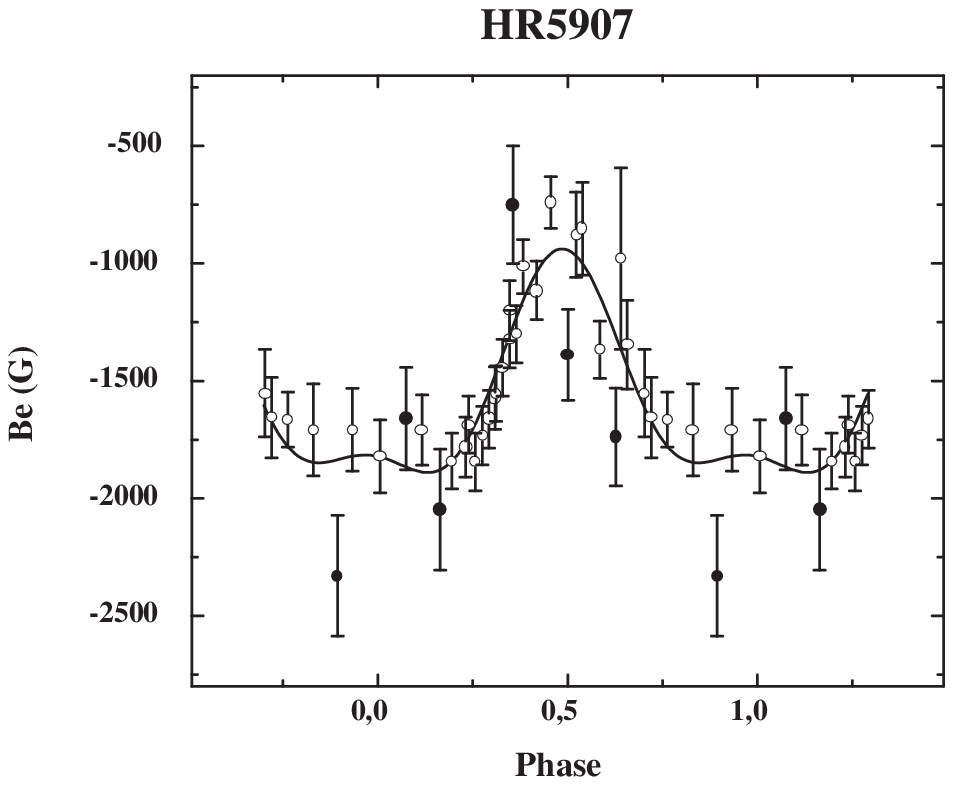}
  \includegraphics[width=4cm]{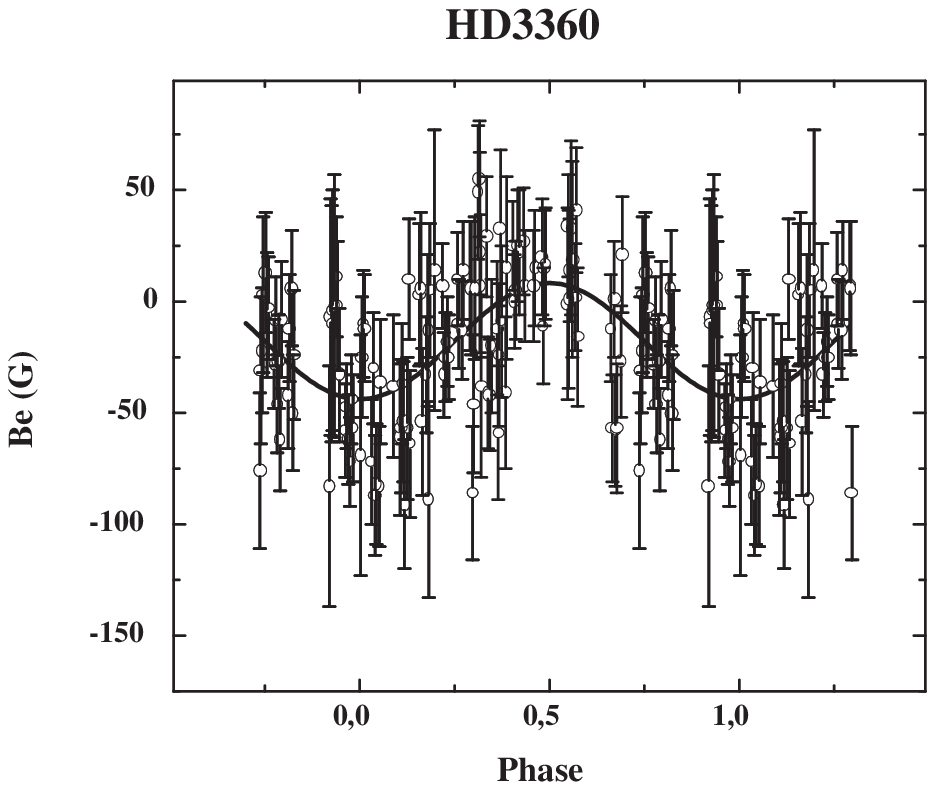}
  \includegraphics[width=4cm]{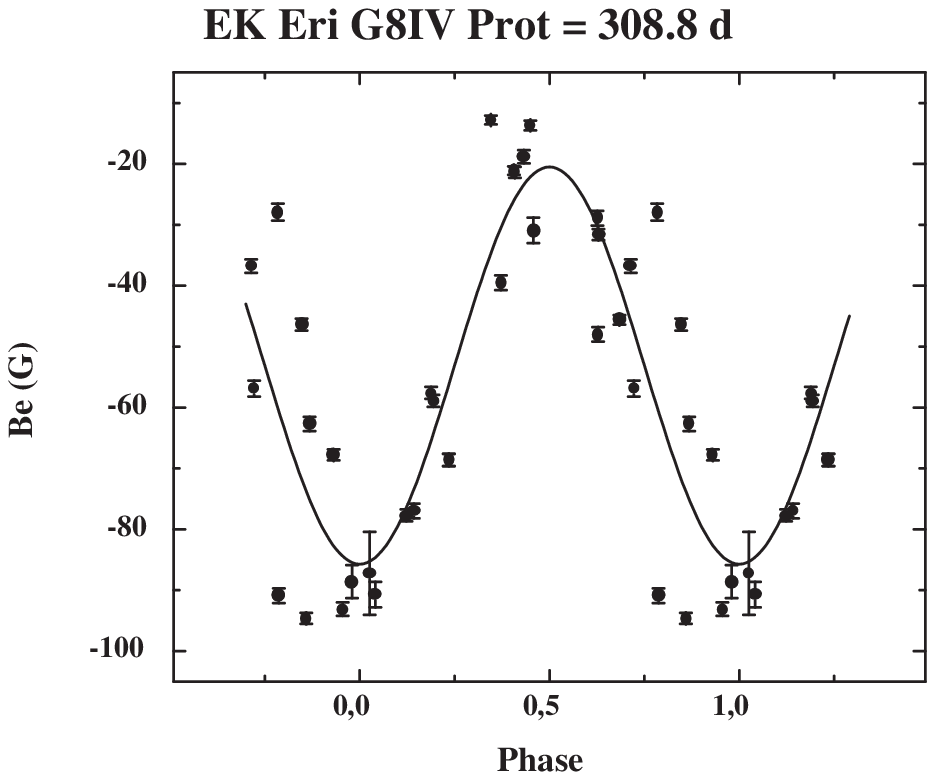}}
\vspace{-5mm}
\caption[]{Magnetic phase curve for the pulsating star HR 5907 ($\beta$ Cep type) with the rotational period
$0.508276$ days determined by Grunhut et al. \cite{Grun_2012} (left panel),
for the pulsating star $\zeta$ Cas = HD 3360 (SPB type) with the rotational period  $5.370447$ days determined by Neiner et al. \cite{Neiner_2003} (center panel), 
for the yellow giant EK Eri -- supposed descendant of a mCP star. Period of rotation equals 308.8 days according to Auriere et al. \cite{Aur_2008},\cite{Aur_2011}. (right panel).}
\label{fig_box2}
\end{center}
\end{figure}

\begin{figure}[!t]
\begin{center}
\hbox{
  \includegraphics[width=4cm]{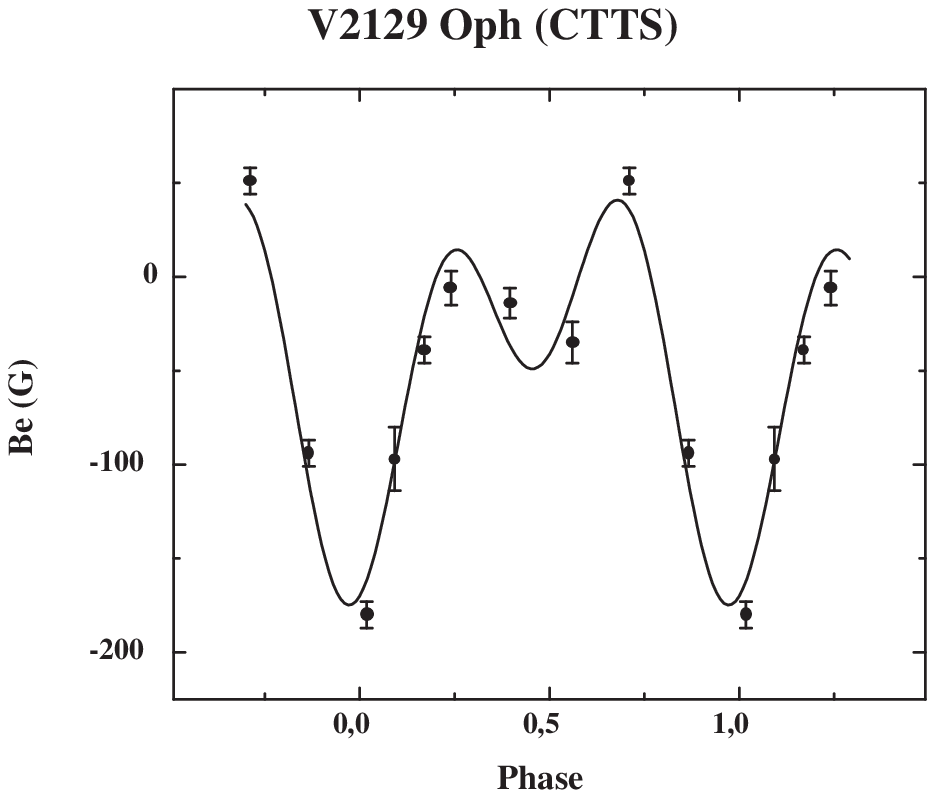}
  \includegraphics[width=4cm]{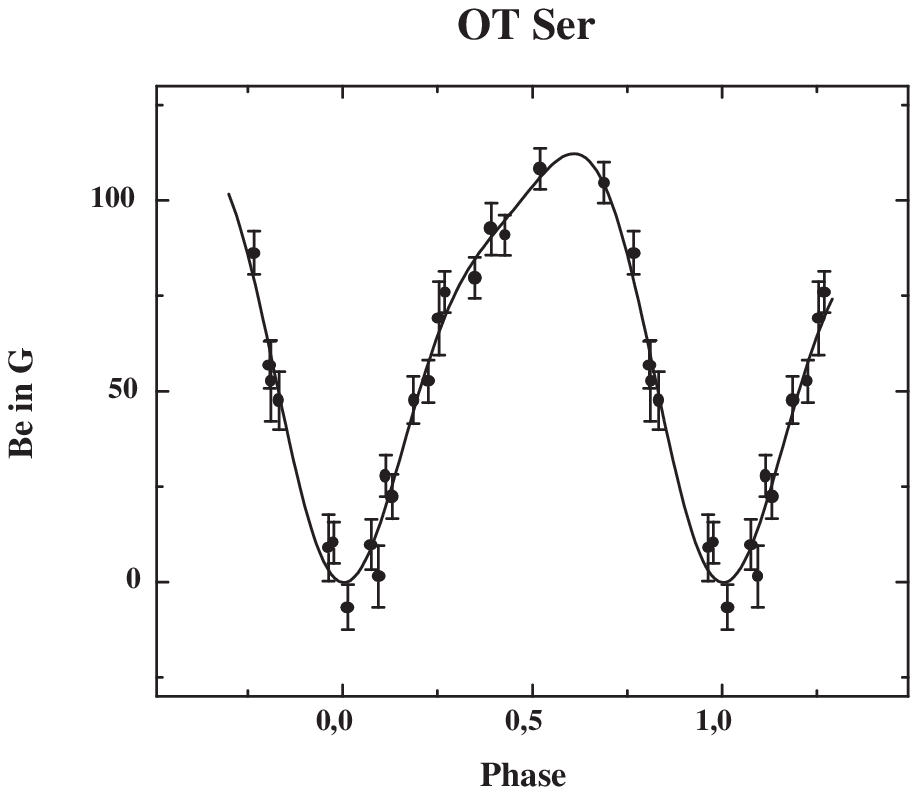}
  \includegraphics[width=4cm]{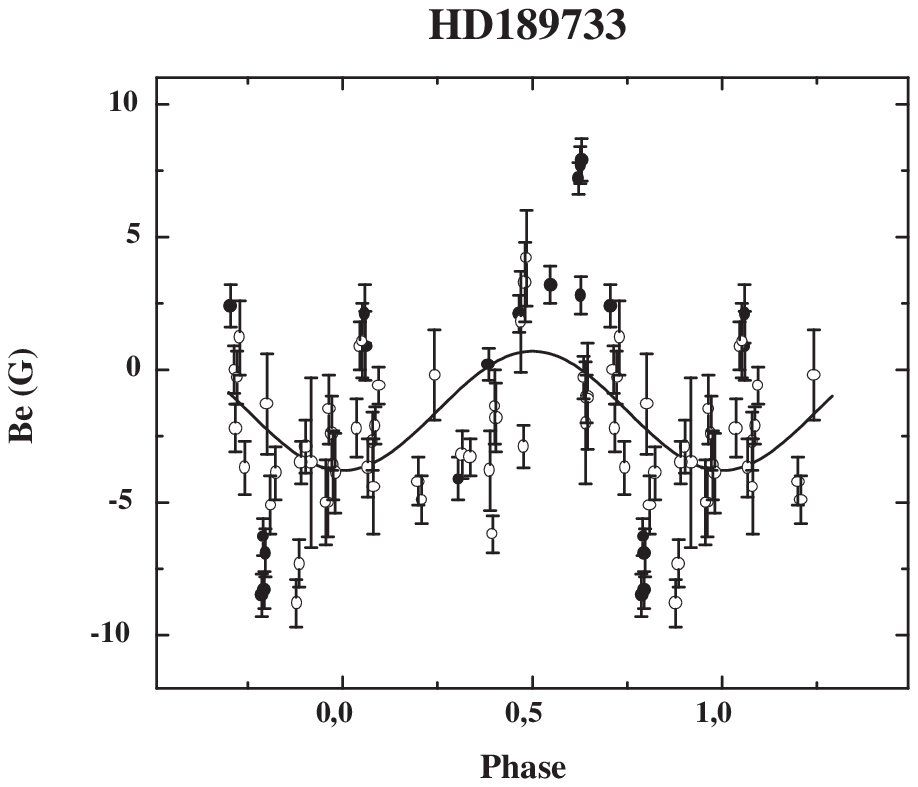}}
\vspace{-5mm}
\caption[]{Magnetic phase curve for $T$ Tau (CTTS) type star V2129 Oph. Period of
   rotation equals 6.53 days according to Donati et al. \cite{Don_2007} (left panel),
for flare star OT Ser with the rotational period 3.424 days (center panel), 
for HD189733 with the rotational period $12^{d}.38$. The star hosts a planetary system (right panel).}
\label{fig_box3}
\end{center}
\end{figure}

\section{Conclusions}
We selected the following, most important conclusions on the magnetic 
activity among stars of various types: \\
1. New class of magnetised objects was recently discovered -- supermassive
hot stars type Of?. Amplitudes of magnetic phase curves (MPC) reach several
hundred G. Of? stars apparently are slow rotators. Configuration of their
magnetic field is represented by an oblique rotator. \\
2. Magnetic fields were found among chemically normal early B stars. MPC's
were obtained for 3 stars of this type. In one object, HD 149438, MPC shows
complicated double wave shape, displayed also by some mCP stars.  \\
3. Magnetic field and its behaviour was best investigated in the group of 
mCP stars. Longitudinal magnetic fields $B_e$ have simple dipole 
configuration in majority of mCP stars (in 86 \% objects). Rotational 
magnetic phase curves often display simple harmonic shape with amplitudes
reaching 10 kG. \\
Remaining 14 \% of investigated mCP stars display more complex MPC
being a superposition of two sine waves and have either dipole 
or more complex structure of their global magnetic fields. Amplitudes 
of rotational $B_e$ variation essentially do not differ from those in 
``sine-wave'' mCP stars. \\
4. Solar-type stars have global magnetic fields of low strength, seldom
approaching few dozens G. Measuring of such low-intensity fields meets
with many methodical difficulties. Therefore, we can only suppose,
that in some investigated stars (in $\xi$ Boo A, for example) MPC's
appear as simple harmonic waves. Very significant 
progress in measuring of magnetic fields in stars was achieved using 
the ZDI method (magnetic cartography of the surface). \\
5. Ae/Be Herbig stars usually exhibit MPC's 
of purely harmonic shape with amplitudes reaching several hundred G. \\
6. MPC's of pulsating $\beta$ Cep stars 
varies with the period of rotation. MPC shows a complicated structure
with low amplitudes of dozens G. Closely related slowly pulsating B stars
(SPB) also display longitudinal magnetic field varying with the period of
rotation. MPC shows a simple harmonic shape with amplitudes reaching several
dozens G. \\
7. T Tau stars have magnetic fields of complex structure, they also
display complex MPC's with amplitudes approaching several
hundred G. Undoubtedly, fields of such a strength have to strongly
influence accretion of matter onto stars. \\
8. Late-type stars -- M dwarfs have global magnetic fields of complex
structure. Magnetic rotational phase curves only roughly can be 
approximated by a superposition of two waves. This was also directly
confirmed by recent observations using the ZDI method. Amplitudes of 
variations of the integrated longitudinal magnetic fields reach
several hundred G. Some stars present an amazing feature, stepwise
creation or anihilation of the global magnetic field and related
$B_e$ variations. \\
9. HD 189733 -- this is typical dwarf of spectral class K2V, where a 
giant planet, ``hot Jupiter'' was found. Central star in the system 
is a solar-like object. The star possesses magnetic field which is 
typical for its spectral class, and its longitudinal component 
varies with the amplitude of several G. \\

In recent years significant progress was attained in the study of stellar
magnetism. While previously one could measure and discuss behaviour of the 
stellar magnetic field only in mCP stars, white dwarfs and the Sun, currently 
we can measure and collect data on the magnetic field for many more types of 
stars ranging from supermassive hot giants to fully convective cold dwarfs of
low mass. One can note significant contribution of the MiMeS collaboration
which has discovered a new class of magnetic objects, supermassive hot giants
Of? type and other magnetised hot stars. \\
Variations of global longitudinal magnetic fields are described by average
curves and define movements of matter close to stars.
It influences physical processes and evolution of stars.  \\

\bigskip
{\it Acknowledgements.}
We acknowledge support from
the Polish Ministry of Science and Higher Education grant No. N N203 511638
and the Russian grant ``Leading Scientific School''  N4308-2012.2.

\end{document}